\documentclass[11pt,floatfix,preprint,showpacs]{revtex4}
\usepackage{epsfig}
\usepackage{graphicx}
\usepackage{dcolumn}
\usepackage{bm}
\usepackage{amsmath}
\usepackage{amssymb}
\usepackage{amsfonts}

\begin{document}

\title{A New Treatment of 2N and 3N Bound States in Three
Dimensions}

\author{W. Gl\"ockle$^1$}

\author{Ch. Elster$^2$}

\author{J. Golak$^3$}

\author{R. Skibi\'nski$^3$}

\author{H. Wita{\l}a$^3$}

\author{H. Kamada$^4$}

\affiliation{$^1$Institut f\"ur theoretische Physik II,
Ruhr-Universit\"at Bochum, D-44780 Bochum, Germany}

\affiliation{$^2$Institute of Nuclear and Particle Physics, Department of Physics and
Astronomy, Ohio University, Athens, OH 45701, USA}

\affiliation{$^3$M. Smoluchowski Institute of Physics, Jagiellonian
University, PL-30059 Krak\'ow, Poland}

\affiliation{$^4$Department of Physics, Faculty of Engineering, Kyushu Institute of
Technology,
 1-1 Sensuicho Tobata, Kitakyushu 804-8550, Japan}

\date{\today}

\begin{abstract}
  The direct treatment of the Faddeev equation for the three-boson system
  in 3 dimensions is generalized to nucleons. The one Faddeev
  equation for identical bosons is replaced by a strictly finite
  set of coupled equations for scalar functions which depend only
  on 3 variables. The spin-momentum dependence  occurring as scalar
  products in 2N and 3N forces accompanied by scalar functions is
  supplemented by a corresponding expansion of the Faddeev
  amplitudes. After removing the spin degrees of freedom by suitable 
operations only
  scalar expressions  depending  on momenta remain. The
  corresponding steps are performed for the deuteron leading to two 
coupled equations.
\end{abstract}


\maketitle \setcounter{page}{1}

\section{Introduction}

With the advent of chiral three-nucleon (3N) forces 
at N$^3$LO \cite{N3LO} which have a rich
spin and momentum structure, the calculation of the 3N system based
on  standard partial wave representations
is rather tedious.  Moreover, for higher angular momenta it is a
challenging task to reliably control  the numerical accuracy~\cite{Hueber}.
For a system of three bosons a numerical treatment working
directly with momentum vectors for bound
and scattering states including model three-boson forces  turned out to be
very feasible~\cite{schadow,liu2,liu3,Lin:2008sy}. Therefore we 
suggest to also formulate
the Faddeev equation for the 3N system directly with vector
variables.
Our proposed formulation is an alternative to Ref.~\cite{Bayegan:2007ih}, where
spin and iso-spin couplings are explicitly carried out.
The key ingredient
   thereby is an operator representation of the $^3$He wave function
   given long time ago by \cite{gerjuoy} and later on modified and strictly
   re-derived from the complete partial wave representation~\cite{fach04}. 
 Thus, the 
   idea is to represent 2N t-operators, the 3N forces
   and the Faddeev components of the 3N wave function as products of
   scalar functions with scalar products of spin operators and
   momentum vectors. Furthermore,  by suitable operations one removes
   the spin operators leading to relations among scalar functions
   of momentum vectors only.

     In Section~\ref{sec_deu} we consider the deuteron as warm-up exercise,
     whereas the central topic, the treatment of the 3N Faddeev
     equation, is given in Section~\ref{sec_bs}. We end with a brief summary
     and outlook in Section~\ref{sec_sum}.

\section{The Deuteron}
\label{sec_deu}

The deuteron momentum wave function has the well known
operator form (see for instance~\cite{fach01})
\begin{eqnarray}
\Psi_{m_d} & = &  \left[ \phi_1(p) + \left(\vec \sigma_{(2)}\cdot \vec p \; 
\vec \sigma_{(3)} \cdot \vec p- \frac{1}{3} p^2\right) \; \phi_2(p) \right]
| 1 m_d\rangle  \cr
& \equiv &  \sum_{ k=1}^2 \phi_k(p)  b_k( \vec \sigma_{(2)}, 
\vec \sigma_{(3)}, 
\vec p) | 1 m_d \rangle,
\end{eqnarray}
where $| 1 m_d \rangle $ describes the state in which two spin $ 1/2 $ states
are coupled to
total  spin 1 and magnetic quantum number $m_d$. The two scalar
functions $ \phi_1(p)$ and $ \phi_2(p)$ are the s- and d-wave
components 
of the deuteron wave function. We label the two nucleons as
particles 2 and 3, and the operator ${\vec
\sigma}_{(i)}$ (i=2,3), represents the corresponding spin operators.

As is well known, rotational-, parity-, and time reversal-invariance 
restrict any NN
potential V to be formed with six terms \cite{wolfenstein}, so that 
the most general NN force has the form
\begin{eqnarray}
V( \vec p,\vec p') = \sum_{j=1}^6 v_j( \vec p,\vec p') \;  w_j(\vec \sigma_{(2)},
\vec \sigma_{(3)}, \vec p,\vec p'),
\end{eqnarray}
with $v_j( \vec p,\vec p')$ being scalar functions. The
spin-momentum scalar operators $w_j (\vec \sigma_{(2)},
\vec \sigma_{(3)}, \vec p,\vec p')$ are given by
\begin{eqnarray}
w_1 & = &  1 \nonumber \\
w_2 & = &  \big( \vec \sigma_{(2)} + \vec \sigma_{(3)}\big) 
\cdot ( \vec p \times \vec p')
\nonumber \\
w_3 & = &  \vec \sigma_{(2)} \cdot ( \vec p \times \vec p') \; \vec \sigma_{(3)}
\cdot ( \vec p \times \vec p') \nonumber \\
w_4 & = &  \vec \sigma_{(2)} \cdot (\vec p + \vec p') \; \vec \sigma_{(3)} \cdot
(\vec p + \vec p') \nonumber \\
w_5 & = & \vec \sigma_{(2)} \cdot (\vec p - \vec p') \; \vec \sigma_{(3)}
\cdot (\vec p - \vec p') \nonumber \\
w_6 & = &  \vec \sigma_{(2)} \cdot (\vec p - \vec p') \; \vec
\sigma_{(3)} \cdot (\vec p + \vec p')\cr 
& + & \vec \sigma_{(2)} \cdot
(\vec p + \vec p') \; \vec \sigma_{(3)} \cdot (\vec p - \vec p'),
\label{eq:3}
\end{eqnarray}
and represent the maximum possible set of invariant operators.
Using these operators the Schr\"odinger equation in integral form reads
\begin{eqnarray}
\Big[\phi_1( p)&+& \left(\vec \sigma_{(2)} \cdot \vec p \; \vec
\sigma_{(3)} \cdot \vec p- \frac{1}{3} p^2 \right) \phi_2(p) \Big] 
| 1 m_d \rangle \cr 
&=&  \frac{1}{ E_d -\frac{p^2}{m}} \int d^3 p' \sum_{j=1}^6 
v_j(\vec p,\vec p')\;  w_j(\vec \sigma_{(2)}, \vec \sigma_{(3)}, \vec p,\vec
p')\cr
 & &  \Big[\phi_1( p') + \left( \vec \sigma_{(2)} \cdot \vec p' \; \vec
\sigma_{(3)} \cdot \vec p' - \frac{1}{3} {p'}^2\right) \phi_2(p') \Big]| 1 m_d
\rangle.
\end{eqnarray}
We  remove the spin dependence by projecting from the left with 
 $ \langle 1 m_d| b_k( \vec \sigma_{(2)}, \vec \sigma_{(3)}, \vec p)$ and
 summing over $ m_d$. This leads to
\begin{eqnarray}
\lefteqn{\sum_{ m_d =- 1}^{1 }  \langle 1 m_d |  b_k( \vec \sigma_{(2)}, \vec
\sigma_{(3)}, \vec p)\sum_{ k'=1}^2 \phi_{k'}(p)  b_{k'}( \vec
\sigma_{(2)} \vec \sigma_{(3)} \vec p)| 1 m_d \rangle}  \cr
 & = & \frac{1}{ E_d -\frac{p^2}{m}} \sum_{ m_d =- 1}^{1 } \int d^3 p' \sum_{j=1}^6
v_j( \vec p,\vec p') w_j(\vec \sigma_{(2)}, \vec \sigma_{(3)}, \vec
p,\vec p')\cr & &  \sum_{ k''=1}^2 \phi_{k''}(p') \;   b_{k''}( \vec
\sigma_{(2)}, \vec \sigma_{(3)}, \vec p')| 1 m_d \rangle  ~.
\label{eq:9}
\end{eqnarray}
If we define the scalar objects
\begin{eqnarray}
A_{ k k'} ( p) & \equiv &  \sum_{ m_d =- 1}^{1 } \langle 1 m_d| b_k( \vec
\sigma_{(2)}, \vec \sigma_{(3)}, \vec p)  b_{k'}( \vec \sigma_{(2)}, \vec
\sigma_{(3)}, \vec p)| 1 m_d \rangle \\
B_{ kjk''} ( \vec p,\vec p') & \equiv&  \sum_{ m_d =- 1}^{1 }  \langle 1
m_d| b_k( \vec \sigma_{(2)}, \vec \sigma_{(3)}, \vec p) w_j( \vec
\sigma_{(2)}, \vec \sigma_{(3)}, \vec p, \vec p')\cr & &  b_{k''}( \vec
\sigma_{(2)}, \vec \sigma_{(3)}, \vec p')| 1 m_d \rangle, 
\end{eqnarray}
which can be calculated once for all we obtain
\begin{eqnarray}
\sum_{ k'=1}^2 A_{k k'}(p) \phi_{k'} (p) =\frac{1}{ E_d
-\frac{p^2}{m}}\int d^3 p' \sum_{j=1}^6 v_j( \vec p,\vec p')\sum_{
k''=1}^2 B_{ kjk''} ( \vec p,\vec p')\phi_{k''}(p') ~.
\end{eqnarray}
This is a set of two coupled equations for the s- and d-wave
components $ \phi_1(p)$ and $\phi_2(p)$.
The summation over $ m_d $ guarantees the scalar nature of 
$A_{ k k'} ( p)$ and $B_{ kjk''} ( \vec p,\vec p')$. 
 Expressions for $A_{ k k'} ( p)$ and $B_{ kjk''} ( \vec
 p,\vec p')$ are given in the Appendix.

\noindent
The azimuthal angle can be trivially integrated out, leading to the
final form of the deuteron equation
\begin{eqnarray}
&&\sum_{ k'=1}^2 A_{ k k'} ( p)\phi_{k'} (p) = \cr
&&\frac{2 \pi}{ E_d
   -\frac{p^2}{m}} 
\sum_{ k''=1}^2 
\int_0^{\infty} dp' {p'}^2\phi_{k''}(p') \int_{-1}^{1} dx
\sum_{j=1}^6 v_j(p,p',x) B_{ kjk''} ( p,p',x),
\end{eqnarray}
 where $x \equiv \hat {\vec p} \cdot \hat {\vec p}'$.

\section{ The 3N bound state}
\label{sec_bs}

The Faddeev equation for the 3N bound state reads
\begin{eqnarray}
 \psi = G_0 t P \psi + ( 1 + G_0 t ) G_0 V^{(1)} ( 1 + P) \psi ~,
\label{faddeev}
 \end{eqnarray}
 where $ G_0$ is the free propagator,  
$t$  the NN $t$-operator for nucleons 2 and 3, the permutation operator
  $P = P_{12}P_{23} + P_{13}P_{23}$.
 The term $V^{(1)}$ is that part of the 3N force, which is symmetrical
 under exchange of nucleons 2 and 3. The choice of the subsystem pair
$2,3$  is of course arbitrary. The Faddeev component is given by $\psi$, 
and the
 total 3N state is then $\Psi =(1 + P) \psi$.

\noindent
We introduce the three possible 3N isospin states
\begin{eqnarray}
| \gamma_0 \rangle & = &  \vert \left( 0 \frac{1}{2}\right) \frac{1}{2}
\rangle \nonumber \\
| \gamma_1 \rangle & = &  | \left( 1 \frac{1}{2}\right)
\frac{1}{2}\rangle \nonumber  \\
| \gamma_2 \rangle & = &  | \left( 1 \frac{1}{2}\right) \frac{ 3}{2}\rangle ,
\end{eqnarray}
with the notation that the 2N subsystem isospin is either 0 or 1 and coupled
with the isospin of the third nucleon to the total isospins
T=1/2 or 3/2.

\noindent
 We expand the Faddeev component $\psi$, the 2N t-matrix $t$,  and $V^{(1)}$
as
\begin{eqnarray}
\psi &=& \sum_{\gamma} | \gamma\rangle \psi_{\gamma} \equiv \sum_{tT} |
\left( t\frac{1}{2}\right) T\rangle \psi_{tT}\\ 
t &=& \sum_{\gamma, \gamma'}|\gamma\rangle t_{\gamma, \gamma'} \langle \gamma'|\\
V^{(1)} &=& \sum_{\gamma, \gamma'}| \gamma \rangle  V_{ \gamma,
\gamma'}^{(1)} \langle \gamma'| ~.
\end{eqnarray}
Then projecting Eq.~(\ref{faddeev}) from the left, we obtain
\begin{eqnarray}
\psi_{\gamma} &=& G_0 \sum_{\gamma', \gamma''} t_{ \gamma,\gamma'}
 \langle \gamma'| P | \gamma''\rangle \psi_{\gamma''}\cr
 &  + &  \sum_{\gamma', \gamma'',
\gamma'''} \langle \gamma| ( 1 + G_0 t) | \gamma' \rangle G_0 V_{ \gamma',
\gamma''}^{(1)} \langle \gamma''| ( 1 + P )| \gamma''' \rangle
 \psi_{\gamma'''} ~.
\label{eq:23}
\end{eqnarray}
We neglect the change of the total 2N isospin and the change  of 
the total 3N isospin  in the 3NF
 and  define
 \begin{eqnarray}
t_{ \gamma,\gamma'} &\equiv& \delta_{t t'} t_{t T T'}\\
V_{ \gamma, \gamma'}^{(1)} &\equiv& \delta_{ T T'} V_{ t t'
T}^{(1)} ~.
\end{eqnarray}

\noindent
Furthermore \cite{Witala:2009ws},
 \begin{eqnarray}
 \langle \gamma| P| \gamma'\rangle = \delta_{TT'} P_{t t'T} = \delta_{TT'}
 F_{ tt' T} ( P_{12}^{sm} P_{23}^{sm}
 + ( -)^{t+t'} P_{13}^{sm} P_{23}^{sm}) ~,
 \end{eqnarray}
 where $F_{ tt' T}$ are geometrical factors and $ P_{ij}^{sm}$
 acts only in spin and momentum space.
  Inserting all that into Eq.~(\ref{eq:23}) yields
\begin{eqnarray}
\psi_{tT} &=& G_0 \sum_{T'} t_{ tTT'} \sum_{t'} F_{t t' T'} \left(
P_{12}^{sm} P_{23}^{sm}
 + ( -)^{t+t'} P_{13}^{sm} P_{23}^{sm}\right) \psi_{t'T'}\cr
 &+& \sum_{T'} \left( \delta_{TT'} + G_0 t_{ t TT'} \right) G_0 \sum_{t'} V_{ t
 t' T'}^{(1)}\cr
 & & \sum_{t''} ( \delta_{ t' t''} + F_{ t' t'' T'} (P_{12}^{sm}
P_{23}^{sm}
 + ( -)^{t'+t''} P_{13}^{sm} P_{23}^{sm})) \psi_{ t'' T'} ~.
\label{eq:27}
 \end{eqnarray}
The charge independence and charge symmetry breaking in the 2N
 force leads to the coupling of total isospin T=1/2 and 3/2.

In the space spanned by the two Jacobi momenta $ \vec p $ and
$ \vec q $ we encounter the off-shell NN $t$-matrix
\begin{eqnarray}
\langle \vec p \vec q| t_{t TT'}| \vec p' \vec q'\rangle = \delta( \vec q -
\vec q') t_{tTT'} ( \vec p, \vec p',E_q)
\end{eqnarray}
with $E_q \equiv E - \frac {3} {4m} q^2$ 
and the matrices for the permutations~\cite{Gl83}
 \begin{eqnarray}
\langle \vec p \vec q| P_{12}^{sm} P_{23}^{sm}| \vec p' \vec q'\rangle & = &
\delta( \vec p - \vec \pi( \vec q, \vec q')) \delta( \vec p' -
\vec \pi'( \vec q, \vec q'))P_{12}^{s} P_{23}^{s} \nonumber \\
\langle \vec p \vec q| P_{13}^{sm} P_{23}^{sm}| \vec p' \vec q'\rangle &=&
\delta( \vec p + \vec \pi( \vec q, \vec q')) \delta( \vec p' +\vec
\pi'( \vec q, \vec q'))P_{13}^{s} P_{23}^{s}
\end{eqnarray}
with
\begin{eqnarray}
\vec \pi( \vec q, \vec q') = \frac{1}{2} \vec q + \vec q'\\
\vec \pi'( \vec q, \vec q')= \vec q + \frac{1}{2} \vec q' ~.
\end{eqnarray}
Consequently with $ \psi_{tT}( \vec p, \vec q) \equiv \langle \vec p\vec
q|\psi_{tT}\rangle $ we get
\begin{eqnarray}
& & \psi_{tT}( \vec p, \vec q) = \frac{1}{ E-\frac{p^2}{m} -
\frac{3}{4m} q^2}  \int d^3 q'\cr
 & & \sum_{T'}\Big( t_{ tTT'} ( \vec p, \vec \pi( \vec q, \vec q'), E_q ) \; 
P_{12}^{s} P_{23}^{s} \; \sum_{t'}F_{ tt' T'} \; 
\psi_{t'T'}( \vec \pi'( \vec q,\vec q'), \vec q') \cr
 &  +  & (-)^{t+t'}t_{ tTT'} ( \vec p, -\vec \pi( \vec q,\vec q'),E_q ) \; 
P_{13}^{s} P_{23}^{s} \; \sum_{t'}F_{ tt' T'} \; 
\psi_{t'T'}( -\vec \pi'( \vec q,\vec q'), \vec q') \Big)\cr
 &  + & \frac{1}{ E-\frac{p^2}{m} - \frac{3}{4m} q^2}
\int d^3 p' d^3 q' \sum_{t'} V_{ tt'T}^{(1)} ( \vec p, \vec q,
\vec p', \vec q') \; \psi_{t' T}( \vec p', \vec q') \cr
 &  + & \frac{1}{ E-\frac{p^2}{m} -
\frac{3}{4m} q^2} \sum_{T'} \int d^3 p' \;  t_{ tTT'} ( \vec p,\vec
p', E_q) \frac{1}{ E-\frac{{p'}^2}{m} - \frac{3}{4m} q^2}\cr
 & & \sum_{t'} \int d^3 p'' d^3 q'' \;  V_{ tt'T'}^{(1)} ( \vec p', \vec q,
\vec p'', \vec q'') \; \psi_{t'T'}( \vec p'', \vec q'')\cr
 &  + & \frac{1}{ E-\frac{p^2}{m} - \frac{3}{4m} q^2}
\int d^3 q' d^3 q'' \cr 
& &\sum_{t'} \Big( V_{ tt'T}^{(1)} ( \vec p,
\vec q, \vec \pi( \vec q', \vec q''), \vec q'') \;  P_{12}^{s}
P_{23}^{s} \; \sum_{t''} F_{ t't'' T} \;  \psi_{t''T}( \vec \pi'( \vec
q',\vec q''), \vec q'')\cr
 & + & (-)^{ t'+ t''}V_{
tt'T}^{(1)} ( \vec p, \vec q,-  \vec \pi( \vec q', \vec q''), \vec
q'')  \;  P_{13}^{s} P_{23}^{s} \; \sum_{t''} F_{ t't'' T} \; \psi_{t''T}(
-\vec \pi'( \vec q',\vec q''), \vec q'') \Big)\cr
 &+& \frac{1}{ E-\frac{p^2}{m} - \frac{3}{4m} q^2} \sum_{T'} 
\int d^3 p' t_{ tTT'} ( \vec p,\vec p', E_q) \frac{1}{
  E-\frac{{p'}^2}{m} 
- \frac{3}{4m}q^2}\cr
& &  \int d^3 q' d^3 q'' \; \sum_{t'} \Big( V_{ tt'T'}^{(1)} ( \vec
p', 
\vec q, \vec
\pi( \vec q',\vec q''), \vec q'') P_{12}^{s} P_{23}^{s} 
  \sum_{t'' } F_{ t't'' T'}\psi_{t''T'}(
\vec \pi'( \vec q', \vec q''), \vec q'')\cr
&+& (-)^{t'+t''}V_{ tt'T'}^{(1)} ( \vec p', \vec q, - \vec \pi( \vec q',
\vec q''), \vec q'') \;  P_{13}^{s} P_{23}^{s}\cr
& & \sum_{t'' } F_{ t't'' T'} \; \psi_{t''T''}( -\vec \pi'(\vec q',\vec q''),
\vec q'') \Big) ~.
\label{eq:33}
\end{eqnarray}

\noindent
The antisymmetry of the Faddeev component under exchange of
particles 2 and 3 imposes the condition
\begin{eqnarray}
(-)^t P_{23}^{sm} \psi_{tT} = \psi_{tT}
 \end{eqnarray}
 which in momentum space reads
\begin{eqnarray}
(-)^t P_{23}^{s} \psi_{tT}( - \vec p,\vec q)  = \psi_{tT}( \vec
p,\vec q) ~.
\label{eq:35}
\end{eqnarray}
 Thus $ \psi_{0T}$  ($\psi_{1T}$)  is even (odd) under exchange  of
 particles 2 and 3 in spin and momenta.

\noindent
This relation can be used to simplify Eq.~(\ref{eq:33}). We regard
\begin{eqnarray}
  P_{13}^s P_{23}^s\psi_{tT}( - \vec p,\vec q) = 
P_{23}^s P_{12}^s P_{23}^s P_{23}^s\psi_{tT}( - \vec p,\vec
  q)=P_{23}^s P_{12}^s P_{23}^s (-)^t\psi_{tT}( \vec p,\vec q)
\end{eqnarray}
which turns Eq.~(\ref{eq:33}) into
\begin{eqnarray}
\lefteqn{ \psi_{tT}( \vec p, \vec q) = } \cr
& & \frac{1}{ E-\frac{p^2}{m} -
\frac{3}{4m} q^2}  \int d^3 q'\cr
& & \sum_{T'} \left( t_{ tTT'} ( \vec p, \vec \pi( \vec q, \vec q'),
E_q )  
+   (-)^t
t_{ tTT'} ( \vec p, -\vec \pi( \vec q,\vec q'),E_q )P_{23}^{s} \right) \cr
 & & \; \;   P_{12}^{s} P_{23}^{s} \sum_{t'}F_{ tt' T'}\psi_{t'T'}(
\vec \pi'( 
\vec q,\vec q'), \vec q')\cr
&+& \frac{1}{ E-\frac{p^2}{m} - \frac{3}{4m} q^2}
\int d^3 p' d^3 q' \sum_{t'} V_{ tt'T}^{(1)} ( \vec p, \vec q,
\vec p', \vec q') \; \psi_{t' T}( \vec p', \vec q') \cr
&+& \frac{1}{ E-\frac{p^2}{m} -
\frac{3}{4m} q^2} \sum_{T'} \int d^3 p' \;  t_{ tTT'} ( \vec p,\vec
p', E_q) \frac{1}{ E-\frac{{p'}^2}{m} - \frac{3}{4m} q^2}\cr
 & & \sum_{t'} \int d^3 p'' d^3 q'' \;  V_{ tt'T'}^{(1)} ( \vec p', \vec q,
\vec p'', \vec q'')\psi_{t'T'}( \vec p'', \vec q'')\cr
 &  + & \frac{1}{ E-\frac{p^2}{m} - \frac{3}{4m} q^2}
\int d^3 q' d^3 q''  \cr
 & &\sum_{t'} \left( V_{ tt'T}^{(1)} ( \vec p, \vec q, \vec \pi( \vec
q', 
\vec q''), \vec q'')
  +  (-)^ {t'} V_{ tt'T}^{(1)} ( \vec p, \vec q, -  \vec \pi( \vec q', 
\vec q''), \vec
q'')P_{23}^s \right) \cr
 & & P_{12}^{s} P_{23}^{s} \; \sum_{t''} F_{ t't'' T} \; \psi_{t''T}(
\vec 
\pi'( \vec q',\vec q''), \vec q'')\cr
&+& \frac{1}{ E- \frac{p^2}{m} - \frac{3}{4m} q^2} \sum_{T'} \int d^3 p' \; 
 t_{ tTT'} ( \vec p,\vec p', E_q) \frac{1}{ E-\frac{{p'}^2}{m} 
- \frac{3}{4m}q^2}\cr
 & &  \int d^3 q' d^3 q''\cr
 & & \sum_{ t'} \left( V_{ tt'T'}^{(1)} ( \vec p', \vec q, \vec 
\pi( \vec q',\vec q''), \vec q'')
    +   (-)^{t'} V_{ tt'T'}^{(1)} ( \vec p', \vec q, 
- \vec \pi( \vec q', \vec q''),
\vec q'') P_{23}^{s} \right) \cr
    & & P_{12}^{s} P_{23}^{s}\sum_{t'' } F_{ t't'' T'} \; 
\psi_{t''T'}( \vec \pi'( \vec q', \vec q''), \vec q'') ~.
\label{eq:37}
\end{eqnarray}

\noindent
We now come to the decisive points. In Ref.~\cite{fach04} we derived the
operator form of the 3N bound state wave function based on the
complete set of partial wave states. Exactly the same arguments
can be applied to the Faddeev components $\psi_{tT}( \vec p, \vec
q)$. Thus, in the notation of Ref.~\cite{fach04} one can write
\begin{eqnarray}
\psi_{tT}( \vec p, \vec q) = \sum_{ i = 1}^ 8 \phi_{ tT}^ {(i)} (
\vec p,\vec q) \; O_i \vert \chi^ m \rangle = \sum_{ i = 1}^ 8 \tilde \phi_{ tT}^
{(i)} ( \vec p,\vec q)| \chi_i \rangle,
\label{eq:38}
\end{eqnarray}
where $|\chi^ m \rangle = | ( 0 \frac{1}{2}) \frac{1}{2} m \rangle $ is a
specific state in which the three spin-$1/2$ states are coupled to the 
total angular momentum
quantum numbers of the 3N bound state. The functions  $\phi_{ tT}^ {(i)} (
\vec p,\vec q)$ are scalar functions and the operators $ O_i$ are
given as
\begin{eqnarray}
O_1 & = &  1 \nonumber \\
O_2 & = &  \vec \sigma(23) \cdot \vec \sigma_{(1)} \nonumber \\
O_3 & = &  \vec \sigma_{(1)} \cdot (\hat{\vec p} \times \hat{\vec q}) \nonumber \\
O_4 & = &  \vec \sigma(23) \cdot \hat{\vec  p} \times \hat{\vec q} \nonumber \\
O_5 & = &  \vec \sigma(23) \cdot  \hat{\vec q} \; \vec \sigma_{(1)} \cdot  
\hat{\vec p} \nonumber \\
O_6 & = &  \vec \sigma(23) \cdot  \hat{\vec p}  \;\vec \sigma_{(1)} \cdot  
\hat{\vec q} \nonumber \\
O_7 & = &  \vec \sigma(23) \cdot  \hat{\vec p}  \;\vec \sigma_{(1)} \cdot  
\hat{\vec p} \nonumber \\
O_8 & = &  \vec \sigma(23) \cdot  \hat{\vec q}  \;\vec \sigma_{(1)} \cdot 
\hat{\vec  q} .
\end{eqnarray}
Here $ \vec \sigma(23) \equiv \frac{1}{2} ( \vec \sigma_{(2)} 
- \vec \sigma_{(3)})$.

\noindent
The second expansion in Eq.~(\ref{eq:38}) is based on
\begin{eqnarray}
\chi_1 & = &  | \chi^ m \rangle  \nonumber \\
\chi_2 & = &  \frac{1}{ \sqrt{3}} \; \vec \sigma( 23) \cdot \vec \sigma_{(1)} |
\chi^ m \rangle  \nonumber \\
\chi_3 & = &   \sqrt{\frac{3}{2}} \frac{1}{i} \; \vec \sigma_{(1)} \cdot \hat{\vec p}
\times \hat{\vec q} | \chi^ m \rangle  \nonumber \\
\chi_4 & = &  \frac{1}{ \sqrt{2}} \Big( i \vec \sigma( 23)  \times
(\hat{\vec p} \times \hat{\vec q})
 - \vec \sigma_{(1)} \times \vec \sigma( 23) \cdot (\hat{\vec p} \times \hat{\vec q}) 
\Big) |
   \chi^ m \rangle  \nonumber \\
\chi_5 & = &  \frac{1}{i} \Big( \vec \sigma( 23) - \frac{ i}{2} \vec
\sigma_{(1)} \times
\vec \sigma( 23)  \Big) \times (\hat{\vec p} \times \hat{\vec q})| \chi^ m \rangle  
\nonumber \\
\chi_6 & = &  \sqrt{\frac{3}{2}} \Big( \vec \sigma( 23) \cdot \hat{\vec p} \;
\vec \sigma_{(1)} \cdot \hat{\vec p}
 - \frac{1}{3} \vec \sigma( 23) \cdot \vec \sigma_{(1)}  \Big) | \chi^
 m 
\rangle  \nonumber \\
\chi_7 & = &   \sqrt{\frac{3}{2}}  \Big(\vec \sigma( 23) \cdot \hat{\vec q} \;
\vec \sigma_{(1)} \cdot \hat{\vec q}
 - \frac{1}{3}\vec \sigma( 23) \cdot \vec \sigma_{(1)}  \Big)| \chi^ m 
\rangle  \nonumber \\
\chi_8 & = &  \frac{3}{2} \frac{ 1}{ \sqrt{5}}  \Big( \vec \sigma( 23)
\cdot \hat{\vec q} \; \vec \sigma_{(1)} \cdot \hat{\vec p} + \vec \sigma( 23) \cdot
\hat{\vec p} \vec \sigma_{(1)} \cdot \hat{\vec q}\cr &  - &  \frac{2}{3} \hat{\vec p}
\cdot \hat{\vec q} \vec \sigma( 23) \cdot \vec \sigma_{(1)} \Big) | \chi^ m
\rangle ~.
\end{eqnarray}
One can express the $|\chi_i \rangle$ in terms of the $ O_i| \chi^ m \rangle $
and vice versa. Furthermore, we use the most general operator structure
of the NN t-matrix $ t_{tTT'}(\vec p,\vec p',E_q)$, which is given by
\begin{eqnarray}
t_{ tTT'}( \vec p, \vec p',E_q) = \sum_{ j=1}^ 6 t_{ t T T'}^ {
(j)} ( \vec p, \vec p',E_q) \; w_j (\vec \sigma_{(2)}, \vec
\sigma_{(3)}, 
\vec p, \vec p').
\label{eq:55}
\end{eqnarray}
Here the matrix elements  $t_{ t T T'}^ { (j)} ( \vec p, \vec p',E_q) $ are scalar
functions and the operators $w_j( \vec \sigma_{(2)}, \vec \sigma_{(3)}, \vec p,
\vec p')$ are  given in Eq.~(\ref{eq:3}).

\noindent
Next,  the 3NF operator $V_{ tt'T}^{(1)} ( \vec p, \vec q, \vec
p', \vec q')$ also allows for an expansion
\begin{eqnarray}
V_{ tt'T}^{(1)} ( \vec p, \vec q, \vec p', \vec q') = \sum_l v_{ t
t' T}^{(l)} (\vec p, \vec q, \vec p', \vec q') \; \Omega_l(\vec
\sigma_{(1)}, \vec \sigma_{(2)}, \vec \sigma_{(3)}, \vec p, \vec q, \vec p',
\vec q')
\label{eq:62}
\end{eqnarray}
with a strictly finite number of terms. In the context of chiral
effective field theory 3NF's have been worked out up to N$^3$LO 
\cite{N3LO,epel02}, where the expressions can be found.
 As an example we provide the operators $ \Omega_l $ for the $ 2 \pi $-exchange
 \cite{coon79,coon81}
\begin{eqnarray}
V^{(1)} &=& \frac{ \vec \sigma_{(2)} \cdot \vec Q \; \vec \sigma_{(3)}
\cdot \vec Q'}{ (Q^2 + {m_{\pi}}^2)({Q'}^2 + {m_{\pi}}^2)} \left( A + B
\vec Q \cdot \vec Q'\right) \vec \tau(2) \cdot \vec \tau(3)\cr 
& + &C \vec \sigma_{(1)} \cdot ( \vec Q \times \vec Q') \; \vec \tau(1) \cdot 
\left( \vec \tau(2) \times \vec \tau(3)\right),
\end{eqnarray}
where $ \vec Q$ and $ \vec Q'$ are momentum transfers of nucleons
2 and 3.

\noindent
Using \cite{coon81}
\begin{eqnarray}
& &  \langle \left(t\frac{1}{2}\right) T |\vec \tau(2) \cdot \vec \tau(3)| 
\left(t' \frac{1}{2}\right) T \rangle = - \delta_{t t'} \; 6 (-)^t
 \left\{{\begin{array}{*{20}c}
   {\frac{1}{2}} & {\frac{1}{2}} & t  \\
   {\frac{1}{2}}  & {\frac{1}{2}} & 1  \\
\end{array}}\right\} \\
& & \langle \left( t\frac{1}{2}\right) T |\vec \tau(1) \cdot ( \vec
 \tau(2) 
\times
\vec \tau(3))|\left(t' \frac{1}{2}\right) T \rangle = i 36 \sqrt{ ( 2
  t+1) 
( 2t' +
1)} (-)^{ t' + \frac{1}{2} +T}\cr & &
\left\{{\begin{array}{*{20}c}
   {\frac{1}{2}} & t' & T  \\
   t  & {\frac{1}{2}} & 1  \\
\end{array}}\right\}
\left\{{\begin{array}{*{20}c}
   1 & 1 & 1  \\
 {\frac{1}{2}}    & {\frac{1}{2}} & t  \\
  {\frac{1}{2}} & {\frac{1}{2}} & t'  \\
\end{array}}\right\},
 \end{eqnarray}
 one obtains
\begin{eqnarray}
V_{ tt'T}^{(1)} ( \vec p, \vec q, \vec p', \vec q') & =& v_{ t t'
T}^{(1)} (\vec p, \vec q, \vec p', \vec q')\cdot {\bf 1} \cr
&+ &  v_{ t t' T}^{(2)} (\vec p, \vec q, \vec p', \vec q') \vec \sigma_{(1)} \cdot
\Big( ( \vec p - \vec p' ) \times ( \vec q - \vec q')\Big),
\end{eqnarray}
with $\vec Q, \vec Q'$ expressed in terms of the Jacobi momenta.
This defines a subset of the $ \Omega_l$ operators with the
accompaning scalar functions  $ v_{ t t' T}^{(l)}$.

\noindent
Now we insert Eqs.~(\ref{eq:38}),(\ref{eq:55}) and (\ref{eq:62})  into
Eq.~(\ref{eq:37}) and obtain
\begin{eqnarray}
\lefteqn{ \sum_{ i = 1}^ 8 \phi_{ tT}^ {(i)} ( \vec p,\vec q) \; O_i ( \vec
\sigma_{(1)}, \vec \sigma_{(2)}, \vec \sigma_{(3)}, \vec p, \vec q) | \chi^
m \rangle = } \cr 
& &\frac{1}{ E-\frac{p^2}{m} - \frac{3}{4m} q^2}  \int d^3 q' \;
  \sum_{T'}\sum_{ j=1}^ 6 \Big( t_{ t T T'}^ { (j)} ( \vec p, \vec \pi
  ( \vec q,
\vec q'), E_q)
 w_j(\vec \sigma_{(2)}, \vec \sigma_{(3)}, \vec p, \vec \pi ( \vec q,\vec q'))\cr
&+&  (-)^t t_{ t T T'}^ { (j)} ( \vec p, - \vec \pi (
\vec q,\vec q'), E_q) w_j( \vec \sigma_{(2)}, \vec \sigma_{(3)}, \vec p,
- \vec \pi ( \vec q,\vec q'))P_{23}^{s} \Big) \cr
& & \sum_{t'}F_{ tt'
T'}\sum_{ k = 1}^ 8 \phi_{ t'T'}^ {(k)} ( \pi'( \vec q,\vec q'),
\vec q') P_{12}^{s} P_{23}^{s} O_k ( \vec \sigma_{(1)}, \vec
\sigma_{(2)}, \vec \sigma_{(3)}, \pi'( \vec q,\vec q'), \vec q') | \chi^
m \rangle \cr
&+& \frac{1}{ E-\frac{p^2}{m} - \frac{3}{4m} q^2}
\int d^3 p' d^3 q' \sum_{t'} \sum_l v_{ t t' T}^{(l)} (\vec p,
\vec q, \vec p', \vec q') \Omega_l(\vec \sigma_{(1)}, \vec \sigma_{(2)},
\vec \sigma_{(3)}, \vec p, \vec q, \vec p', \vec q') \cr
& & \sum_{ k = 1}^ 8 \phi_{ t'T}^ {(k)} ( \vec p',\vec q') O_k (\vec \sigma_{(1)},
\vec \sigma_{(2)}, \vec \sigma_{(3)}, \vec p', \vec q') | \chi^ m \rangle \cr
&+& \frac{1}{ E-\frac{p^2}{m} -
\frac{3}{4m} q^2} \sum_{T'} \int d^3 p' \sum_{ j=1}^ 6 (
t_{ t T T'}^ { (j)} ( \vec p, \vec p' , E_q)
\;  w_j(\vec \sigma_{(2)}, \vec \sigma_{(3)}, \vec p, \vec p') \cr
& & \frac{1}{ E-\frac{{p'}^2}{m} - \frac{3}{4m} q^2}
  \sum_{t'} \int d^3 p'' d^3 q''
\sum_l v_{ t t' T'}^{(l)} (\vec p', \vec q, \vec p'', \vec q'')
\Omega_l(\vec \sigma_{(1)}, \vec \sigma_{(2)}, \vec \sigma_{(3)}, \vec p',
\vec q, \vec p'', \vec q'')\cr & &  \sum_{ k = 1}^ 8 \phi_{ t'T'}^
{(k)} ( \vec p'',\vec q'') O_k (\vec \sigma_{(1)}, \vec \sigma_{(2)},
\vec \sigma_{(3)}, \vec p'', \vec q'') | \chi^ m \rangle \cr
 &+& \frac{1}{ E-\frac{p^2}{m} - \frac{3}{4m} q^2}
\int d^3 q' d^3 q'' \sum_{t'}\cr
& & \Big( \sum_l v_{ t t' T}^{(l)} (\vec p, \vec q,  \vec \pi(\vec q', 
\vec q''), \vec q'')
 \Omega_l(\vec \sigma_{(1)}, \vec \sigma_{(2)}, \vec \sigma_{(3)},
 \vec p, 
\vec q,\vec
\pi(\vec q', \vec q''), \vec q'')\cr & &   +  (-)^ {t'}\sum_l v_{
t t' T}^{(l)} (\vec p, \vec q, - \vec \pi( \vec q', \vec q''),
\vec q'') \Omega_l(\vec \sigma_{(1)}, \vec \sigma_{(2)}, \vec
\sigma_{(3)}, \vec p, \vec q, - \vec \pi( \vec q', \vec q''), \vec
q'') P_{23}^s \Big) \cr
 & &  \sum_{t''} F_{ t't'' T}\sum_{ k = 1}^ 8 \phi_{
t''T}^ {(k)} ( \vec \pi'( \vec q',\vec q''), \vec q'') P_{12}^{s}
P_{23}^{s}O_k ( \vec \sigma_{(1)}, \vec \sigma_{(2)}, \vec \sigma_{(3)},
\pi'( \vec q',\vec q''), \vec q'') | \chi^ m \rangle \cr
 &+& \frac{1}{ E-\frac{p^2}{m} - \frac{3}{4m} q^2} \sum_{T'} \int d^3 p'
\sum_{ j=1}^ 6 ( t_{ t T T'}^ { (j)} ( \vec p, \vec p' , E_q)
 w_j(\vec \sigma_{(2)}, \vec \sigma_{(3)}, \vec p, \vec p')\cr
& &    \frac{1}{ E-\frac{{p'}^2}{m} - \frac{3}{4m}q^2}  \int d^3
q' d^3 q''\cr
 & & \sum_{ t'} \Big( \sum_l v_{ t t' T'}^{(l)} (\vec p', \vec q,  
\vec \pi( \vec q', \vec q''), \vec
q'') \Omega_l(\vec \sigma_{(1)}, \vec \sigma_{(2)}, \vec \sigma_{(3)}, \vec
p', \vec q, \vec \pi( \vec q', \vec q''), \vec q'')\cr
 &+& (-)^{t'}\sum_l v_{ t t' T'}^{(l)} (\vec p', \vec q,  -\vec \pi(
\vec q', \vec q''), \vec q'') \Omega_l(\vec \sigma_{(1)}, \vec
\sigma_{(2)}, \vec \sigma_{(3)}, \vec p', \vec q, - \vec \pi( \vec q',
\vec q''), \vec q'') P_{23}^{s} \Big)\cr 
& & \sum_{t''}F_{ t't'' T'}
\sum_{ k = 1}^ 8 \phi_{ t''T'}^ {(k)} ( \vec \pi'( \vec q',\vec q''),
\vec q'')P_{12}^{s} P_{23}^{s} O_k ( \vec \sigma_{(1)}, \vec
\sigma_{(2)}, \vec \sigma_{(3)}, \pi'( \vec q',\vec q''), \vec q'') |
\chi^ m \rangle.
\label{eq:67}
\end{eqnarray}

\noindent
This can be further simplified by using
\begin{eqnarray}
\lefteqn{ P_{12}^{s} P_{23}^{s} \; O_k ( \vec \sigma_{(1)}, \vec \sigma_{(2)},
\vec \sigma_{(3)}, \vec p, \vec q ) | \chi^ m \rangle   = }\cr
& &  - \frac{1}{2}
O_k ( \vec \sigma_{(2)}, \vec \sigma_{(3)}, \vec \sigma_{(1)}, \vec p,\vec q)
( 1 + \vec \sigma(23)\cdot \vec \sigma_{(1)}) |\chi^ m \rangle ,
\label{eq:68}
 \end{eqnarray}
 which turns Eq.~(\ref{eq:67}) into
\begin{eqnarray}
\lefteqn{ \sum_{ i = 1}^ 8 \phi_{ tT}^ {(i)} ( \vec p,\vec q) O_i ( \vec
\sigma_{(1)}, \vec \sigma_{(2)}, \vec \sigma_{(3)}, \vec p, \vec q) | \chi^
m \rangle = } \cr
& & \frac{1}{ E-\frac{p^2}{m} - \frac{3}{4m} q^2}  \int d^3 q' \;
\sum_{T'} \sum_{ j=1}^ 6 \Big( t_{ t T T'}^ { (j)} ( \vec p, \vec \pi
( 
\vec q,\vec q'), E_q)
 w_j(\vec \sigma_{(2)}, \vec \sigma_{(3)}, \vec p, \vec \pi ( \vec q,\vec
 q'))\cr
& + &  (-)^t t_{ t T T'}^ { (j)} ( \vec p, - \vec \pi ( \vec
q,\vec q'), E_q) w_j( \vec \sigma_{(2)}, \vec \sigma_{(3)}, \vec p, -
\vec \pi ( \vec q,\vec q'))P_{23}^{s} \Big) \cr
 & & \sum_{t'}F_{ tt'
T'}\sum_{ k = 1}^ 8 \phi_{ t'T'}^ {(k)} ( \pi'( \vec q,\vec q'),
\vec q') (-\frac{1}{2}) O_k ( \vec \sigma_{(2)}, \vec \sigma_{(3)}, \vec
\sigma_{(1)}, \pi'( \vec q,\vec q'), \vec q')\cr & &  ( 1 + \vec
\sigma(23)\cdot \vec \sigma_{(1)}) |\chi^ m \rangle \cr
&+& \frac{1}{ E-\frac{p^2}{m} - \frac{3}{4m} q^2}
\int d^3 p' d^3 q' \sum_{t'} \sum_l v_{ t t' T}^{(l)} (\vec p,
\vec q, \vec p', \vec q') \Omega_l(\vec \sigma_{(1)}, \vec \sigma_{(2)},
\vec \sigma_{(3)}, \vec p, \vec q, \vec p', \vec q') \cr
& & \sum_{ k = 1}^ 8 \phi_{ t'T}^ {(k)} ( \vec p',\vec q') O_k (\vec \sigma_{(1)},
\vec \sigma_{(2)}, \vec \sigma_{(3)}, \vec p', \vec q') | \chi^ m \rangle \cr
&+& \frac{1}{ E-\frac{p^2}{m} -
\frac{3}{4m} q^2} \sum_{T'} \int d^3 p' \sum_{ j=1}^ 6  t_{ t T
T'}^ { (j)} ( \vec p, \vec p' , E_q)
 w_j(\vec \sigma_{(2)}, \vec \sigma_{(3)}, \vec p, \vec p')
 \frac{1}{ E-\frac{{p'}^2}{m} - \frac{3}{4m} q^2}\cr
 & & \sum_{t'} \int d^3 p'' d^3 q''
\sum_l v_{ t t' T'}^{(l)} (\vec p', \vec q, \vec p'', \vec q'')
\Omega_l(\vec \sigma_{(1)}, \vec \sigma_{(2)}, \vec \sigma_{(3)}, \vec p',
\vec q, \vec p'', \vec q'')\cr & &  \sum_{ k = 1}^ 8 \phi_{ t'T'}^
{(k)} ( \vec p'',\vec q'') O_k (\vec \sigma_{(1)}, \vec \sigma_{(2)},
\vec \sigma_{(3)}, \vec p'', \vec q'') | \chi^ m \rangle \cr
&+& \frac{1}{ E-\frac{p^2}{m} - \frac{3}{4m} q^2}
\int d^3 q' d^3 q'' \sum_{t'} \cr
& & \Big( \sum_l v_{ t t' T}^{(l)} (\vec p, \vec q,  \vec \pi(\vec q', 
\vec q''), \vec q'')
 \Omega_l(\vec \sigma_{(1)}, \vec \sigma_{(2)}, \vec \sigma_{(3)}, 
\vec p, \vec q,\vec
\pi(\vec q', \vec q''), \vec q'')\cr & &   +  (-)^ {t'}\sum_l v_{
t t' T}^{(l)} (\vec p, \vec q, - \vec \pi( \vec q', \vec q''),
\vec q'') \Omega_l(\vec \sigma_{(1)}, \vec \sigma_{(2)}, \vec
\sigma_{(3)}, \vec p, \vec q, - \vec \pi( \vec q', \vec q''), \vec
q'') P_{23}^s \Big) \cr
& & \sum_{t''} F_{ t't'' T} \sum_{ k = 1}^ 8 \phi_{ t''T}^ {(k)} (
\vec \pi'( 
\vec q',\vec
q''), \vec q'') (-\frac{1}{2}) O_k ( \vec \sigma_{(2)}, \vec
\sigma_{(3)}, \vec \sigma_{(1)}, \vec \pi'( \vec q',\vec q''), \vec
q'')\cr 
& &  ( 1 + \vec \sigma(23)\cdot \vec \sigma_{(1)}) |\chi^
m \rangle \cr
 &+& \frac{1}{ E-\frac{p^2}{m} - \frac{3}{4m} q^2} \sum_{T'} \int d^3 p'
\sum_{ j=1}^ 6  t_{ t T T'}^ { (j)} ( \vec p, \vec p' , E_q)
 w_j(\vec \sigma_{(2)}, \vec \sigma_{(3)}, \vec p, \vec p')\cr
& &    \frac{1}{ E-\frac{{p'}^2}{m} - \frac{3}{4m}q^2}  \int d^3
q' d^3 q''\cr
& & \sum_{ t'} \Big(\sum_l v_{ t t' T'}^{(l)} (\vec p', \vec q,  \vec \pi( \vec
q', \vec q''), \vec q'') \Omega_l(\vec \sigma_{(1)}, \vec \sigma_{(2)},
\vec \sigma_{(3)}, \vec p', \vec q, \vec \pi( \vec q', \vec q''), \vec
q'')\cr
 &+& (-)^{t'}\sum_l v_{ t t' T'}^{(l)} (\vec p', \vec q,  -\vec \pi(
\vec q', \vec q''), \vec q'') \Omega_l(\vec \sigma_{(1)}, \vec
\sigma_{(2)} \vec \sigma_{(3)} \vec p', \vec q, - \vec \pi( \vec q',
\vec q''), \vec q'') P_{23}^{s} \Big) \cr
 & &  \sum_{t''
} F_{ t't'' T'}\sum_{ k = 1}^ 8 \phi_{ t''T'}^ {(k)} ( \vec \pi'(
\vec q',\vec q''), \vec q'')(-\frac{1}{2}) O_k ( \vec
\sigma_{(2)}, \vec \sigma_{(3)}, \vec \sigma_{(1)}, \vec \pi'( \vec q',\vec
q'' ),\vec q'')\cr & &  ( 1 + \vec \sigma(23)\cdot \vec
\sigma_{(1)}) |\chi^ m \rangle.
\label{eq:69}
\end{eqnarray}

\noindent
Analogous to Eq.~(\ref{eq:9}) in case of the deuteron we now project from the
left with $ \langle \chi^ m|  O_k ( \vec \sigma_{(1)}, \vec \sigma_{(2)}, \vec
\sigma_{(3)}, \vec p, \vec q) $ and sum over $m$. This leads to the
following definitions, which are all scalars
\begin{equation}
C_{ki}( \vec p \vec q,\vec p \vec q )  = \sum_{m}  \langle \chi^ m|  O_k ( \vec
\sigma_{(1)}, \vec \sigma_{(2)}, \vec \sigma_{(3)}, \vec p, \vec q) 
 O_i ( \vec \sigma_{(1)}, \vec \sigma_{(2)}, \vec \sigma_{(3)}, \vec p, \vec
q)|\chi^ m \rangle \\
\label{eq42}
\end{equation} 
\begin{eqnarray}
\lefteqn{D_{kjk'} (\vec p \vec q,  \vec p' 
\vec q',  \vec p'' \vec q'') = }\cr
& & -\frac{1}{2} \sum_{m} \langle \chi^ m|  O_k ( \vec \sigma_{(1)}, \vec
\sigma_{(2)}, \vec \sigma_{(3)}, \vec p, \vec q)  w_j(\vec \sigma_{(2)},
\vec \sigma_{(3)}, \vec p', \vec q')\cr
 & & O_{k'} ( \vec \sigma_{(2)}, \vec
\sigma_{(3)}, \vec \sigma_{(1)}, \vec p'', \vec q'') ( 1 + \vec \sigma(
23)\cdot \vec \sigma_{(1)} ) | \chi^ m \rangle 
\label{eq43}
\end{eqnarray}
\begin{eqnarray}
\lefteqn{D_{kjk'}^{'} (\vec p \vec q,  \vec p' \vec q',  \vec p'' \vec
q'') = } \cr
 & & \frac{1}{2} \sum_{m} \langle \chi^ m|  O_k ( \vec \sigma_{(1)},
\vec \sigma_{(2)}, \vec \sigma_{(3)}, \vec p, \vec q)  w_j(\vec
\sigma_{(2)}, \vec \sigma_{(3)}, \vec p', \vec q')\cr
 & & O_{k'} ( \vec \sigma_{(3)} ,\vec
\sigma_{(2)}, \vec \sigma_{(1)}, \vec p'', \vec q'') ( 1 - \vec \sigma(
23)\cdot \vec \sigma_{(1)} )| \chi^ m \rangle 
\label{eq44}
\end{eqnarray}
\begin{eqnarray}
\lefteqn{E_{klk'} (\vec p \vec q , \vec p \vec q  \vec p' \vec q', 
\vec p' \vec q') = } \cr
& & \sum_{m} \langle \chi^
m| O_k ( \vec \sigma_{(1)}, \vec \sigma_{(2)}, \vec \sigma_{(3)}, \vec
p, 
\vec q) \\ 
& & \Omega_l(\vec \sigma_{(1)}, \vec \sigma_{(2)}, \vec
\sigma_{(3)}, \vec p, \vec q, \vec p', \vec q')
\;  O_{k'} ( \vec
\sigma_{(1)}, \vec \sigma_{(2)}, \vec \sigma_{(3)}, \vec p', \vec
q')|\chi^ m 
\rangle
\nonumber 
\label{eq45}
\end{eqnarray}
\begin{eqnarray}
\lefteqn{F_{kj lk'} (\vec p \vec q,  \vec p \vec p',  \vec p' \vec q \vec
p'' \vec q'', \vec p'' \vec q'') = } \cr
 & & \sum_{m}  \langle \chi^ m|  O_k
( \vec \sigma_{(1)}, \vec \sigma_{(2)}, \vec \sigma_{(3)}, \vec p, \vec q)
w_j(\vec \sigma_{(2)}, \vec \sigma_{(3)}, \vec p, \vec p')\cr 
& & \Omega_l(\vec \sigma_{(1)}, \vec \sigma_{(2)}, \vec \sigma_{(3)}, \vec p',
\vec q, \vec p'',\vec q'') 
O_{k'} ( \vec \sigma_{(2)}, \vec \sigma_{(3)}, \vec \sigma_{(1)}, \vec
p'', 
\vec q'')\cr
& &  ( 1 + \vec \sigma(23) \cdot \vec \sigma_{(1)})| \chi^ m \rangle 
\label{eq46}
\end{eqnarray}
\begin{eqnarray}
\lefteqn{G_{klk'}  (\vec p \vec q ,\vec p \vec q \vec p' \vec q', \vec p''
\vec q'') = } \cr
& & -\frac{1}{2}\sum_{m}  \langle \chi^ m|  O_k ( \vec
\sigma_{(1)}, \vec \sigma_{(2)}, \vec \sigma_{(3)}, \vec p, \vec q)
\Omega_l(\vec \sigma_{(1)}, \vec \sigma_{(2)}, \vec \sigma_{(3)}, \vec p,
\vec q, \vec p', \vec q')\cr 
& &
 O_{k'} ( \vec \sigma_{(2)}, \vec \sigma_{(3)}, \vec \sigma_{(1)}, \vec
p'', \vec q'') \;
  ( 1 + \vec \sigma(23) \cdot \vec \sigma_{(1)})| \chi^ m \rangle 
\label{eq47}
\end{eqnarray}
\begin{eqnarray}
\lefteqn{G_{klk'}^{'}  (\vec p \vec q ,\vec p \vec q \vec p' \vec q', \vec
p'' \vec q'') = }\cr
&  & \frac{1}{2} \sum_{m}  \langle \chi^ m|  O_k ( \vec
\sigma_{(1)}, \vec \sigma_{(2)}, \vec \sigma_{(3)}, \vec p, \vec q)
\Omega_l(\vec \sigma_{(1)}, \vec \sigma_{(2)}, \vec \sigma_{(3)}, \vec p,
\vec q, \vec p', \vec q')\cr 
& &  O_{k'} ( \vec \sigma_{(3)}, \vec
\sigma_{(2)} \vec \sigma_{(1)} \vec p'', \vec q'')\; 
  ( 1 - \vec \sigma(23) \cdot \vec \sigma_{(1)})| \chi^ m \rangle 
\label{eq48}
\end{eqnarray}
\begin{eqnarray}
\lefteqn{H_{kjlk'} (\vec p \vec q,  \vec p \vec p', \vec p' \vec q  \vec
p'' \vec q'', \vec p''' \vec q'') = } \cr
& & -\frac{1}{2}\sum_{m} \langle \chi^ m|  
O_k ( \vec \sigma_{(1)}, \vec \sigma_{(2)}, \vec \sigma_{(3)}, \vec p,
\vec q)w_j(\vec \sigma_{(2)}, \vec \sigma_{(3)}, \vec p, \vec p')\cr 
& &
\Omega_l(\vec \sigma_{(1)}, \vec \sigma_{(2)}, \vec \sigma_{(3)}, \vec p',
\vec q, \vec p'', \vec q'') 
 O_{k'} ( \vec \sigma_{(2)}, \vec
\sigma_{(3)}, \vec \sigma_{(1)}, \vec p''', \vec q'')\cr
& &  ( 1 + \vec \sigma(23) \cdot \vec \sigma_{(1)})| \chi^ m \rangle 
\label{eq49}
\end{eqnarray}
\begin{eqnarray}
\lefteqn{ H_{kjlk'}^{'} (\vec p \vec q, \vec p \vec p', \vec p' \vec q
\vec p'' \vec q'', \vec p''' \vec q'') = } \cr
& & 
 \frac{1}{2} \sum_{m}\langle 
\chi^ m| O_k ( \vec \sigma_{(1)}, \vec \sigma_{(2)}, \vec \sigma_{(3)},
\vec p, \vec q)w_j(\vec \sigma_{(2)}, \vec \sigma_{(3)}, \vec p, \vec p')\cr 
& & \Omega_l(\vec \sigma_{(1)}, \vec \sigma_{(2)}, \vec \sigma_{(3)},
\vec p',
\vec q,\vec
p'',\vec q'') 
 O_{k'} ( \vec \sigma_{(3)}, \vec \sigma_{(2)}, \vec \sigma_{(1)} \vec
 p''', 
\vec q'')\cr
& &  ( 1 - \vec \sigma(23) \cdot \vec \sigma_{(1)})| \chi^ m \rangle  ~.
\label{eq50}
\end{eqnarray}

\noindent
Using these expressions we end up with the final form of the Faddeev equation
\begin{eqnarray}
\lefteqn{\sum_{ i = 1}^ 8 C_{ki}(\vec p \vec q,\vec p \vec q )  \; 
\phi_{ tT}^ {(i)} ( \vec
p,\vec q)  =} \cr
& & \frac{1}{ E-\frac{p^2}{m} - \frac{3}{4m} q^2} \; \int
d^3 q' \; \sum_{T'} \sum_{ k' = 1}^ 8 \cr 
& & 
\sum_{ j=1}^ 6 \Big( 
t_{t T T'}^ { (j)} ( \vec p, \vec \pi ( \vec q,\vec q'), E_q) D_{kjk'}
(\vec p \vec q, \vec p \vec \pi( \vec q,\vec q'), \vec \pi'( \vec
q,\vec q') \vec q') \cr
&+&  (-)^t t_{ t T T'}^ { (j)} ( \vec p, - \vec \pi ( \vec
q,\vec q'), E_q)    D_{kjk'}^{'} (\vec p \vec q, \vec p (-\vec
\pi( \vec q,\vec q')), \vec \pi'( \vec q,\vec q') \vec q')\Big) \cr 
& &  \sum_{t'}F_{ tt' T'}\phi_{ t'T'}^ {(k')} ( \pi'( \vec q,\vec
q') \vec q') \cr
&+& \frac{1}{ E-\frac{p^2}{m} - \frac{3}{4m} q^2}
\int d^3 p' d^3 q' \sum_{t'} \sum_l v_{ t t' T}^{(l)} (\vec p,
\vec q, \vec p', \vec q') \sum_{ k' = 1}^ 8 E_{klk'}( \vec p \vec
q, \vec p \vec q \vec p' \vec q', \vec p' \vec q')\phi_{t'T}^ {(k')} 
( \vec p',\vec q')\cr
&+& \frac{1}{ E-\frac{p^2}{m} -
\frac{3}{4m} q^2} \sum_{T'} \int d^3 p' \sum_{ j=1}^ 6  t_{ t T
T'}^ { (j)} ( \vec p, \vec p' , E_q)
 \frac{1}{ E-\frac{{p'}^2}{m} - \frac{3}{4m} q^2}\cr
& & \sum_{t'} \int d^3 p'' d^3 q''
\sum_l v_{ t t' T'}^{(l)} (\vec p', \vec q, \vec p'', \vec q'')
\sum_{ k' = 1}^ 8 F_{kjlk'} ( \vec p \vec q,\vec p \vec p', \vec
p' \vec q \vec p'' \vec q'',\vec p'' \vec q'')\phi_{ t'T'}^ {(k')}
( \vec p'',\vec q'') \cr
&+& \frac{1}{ E-\frac{p^2}{m} -
\frac{3}{4m} q^2} \int d^3 q' d^3 q'' \sum_{t'} \cr
& & \Big( \sum_l v_{ t t' T}^{(l)} (\vec p, \vec q,  \vec \pi(\vec q', 
\vec q''), \vec q'')
\sum_{ k' = 1}^ 8 G_{klk'} ( \vec p \vec q, \vec p \vec q \vec
\pi(\vec q', \vec q'') \vec q'', \vec \pi'(\vec q', \vec q'')
\vec q'')  \cr
 & + &   (-)^ {t'}\sum_l v_{ t t' T}^{(l)} (\vec p, \vec q, - \vec
\pi( 
\vec q', \vec q''),
\vec q'') \sum_{ k' = 1}^ 8 G_{klk'}^{'} ( \vec p \vec q, \vec p
\vec q (-\vec \pi(\vec q', \vec q'')) \vec q'', \vec \pi'(\vec
q', \vec q'') \vec q'') \Big)\cr 
& & \sum_{t''} F_{ t't'' T} \phi_{
t''T}^ {(k')} ( \vec \pi'( \vec q',\vec q''), \vec q'')\cr
&+& \frac{1}{ E-\frac{p^2}{m} - \frac{3}{4m} q^2} \sum_{T'} \int d^3 p'
\sum_{ j=1}^ 6  t_{ t T T'}^ { (j)} ( \vec p, \vec p' , E_q)
\frac{1}{ E-\frac{{p'}^2}{m} - \frac{3}{4m}q^2}  \int d^3 q' d^3
q''\cr
& & \sum_{ t'} \Big(\sum_l v_{ t t' T'}^{(l)} (\vec p', \vec q,  \vec
\pi( 
\vec q', \vec q''), \vec q'')
 \sum_{ k' = 1}^ 8 H_{kjlk'} ( \vec p \vec q, \vec p \vec p', \vec p'
 \vec q 
\vec \pi(\vec q', \vec q'')\vec q'',\vec \pi'(\vec q', \vec q'')\vec
 q'')\cr
&+& (-)^{t'}\sum_l v_{ t t' T'}^{(l)} (\vec p', \vec q,  -\vec \pi( \vec q',
\vec q''), \vec q'')  \cr
& & \sum_{ k' = 1}^ 8 H_{kjlk'}^{'} ( \vec p \vec q, \vec p \vec p',
\vec p' 
\vec q
(-\vec \pi(\vec q', \vec q''))\vec q'',\vec \pi'(\vec q', \vec
q'')\vec q'') \Big) \cr
& & \sum_{t'' } F_{ t't'' T'}  \phi_{ t''T'}^ {(k')} ( \vec \pi'(
\vec q',\vec q''), \vec q'').
\label{eq:79}
\end{eqnarray}

The first part of Eq.~(\ref{eq:79}) refers to NN forces only and
provides 
the bulk of
the binding energy. The remaining parts refer to the action of the
3NF and its interplay with NN forces.

It should be stressed, that the momenta only
occur in  scalar expressions (\ref{eq42})-(\ref{eq50})  which has to be
determined only once. For selected examples we refer to the Appendix.

Due to the expansion of Eq.~(\ref{eq:38}) the number of coupled equations 
to solve is
strictly finite. There at most three 3N isospin states and therefore the
number of equations is
$N= 3\times 8 = 24$. If one neglects charge independence and
charge symmetry breaking in the 2N system then $N=8$ (see also
\cite{Bayegan:2007ih}).  
Moreover as
has been documented in \cite{fach04,kotlyar} only very few out of the 8
components are dominant in the expansion of the full $^3$He state.
We expect that a similar reduction takes place also for the
Faddeev amplitude.

If one neglects the 3NF only the first part in Eq.~(\ref{eq:79}) is present.
The big question at present,  however, is to  investigate the
contribution of the N$^3$LO 3NF's , which are parameter free and
therefore the full set has to be treated.

The final remark refers to symmetry
requirements for the unknown scalar amplitudes $\phi_{tT}^{ (i)} (
\vec p, \vec q)$. They follow from Eq.~(\ref{eq:35}) and the expansion given in
Eq.~(\ref{eq:38}):
\begin{eqnarray}
 P_{23}^ s \; \psi_{tT}( -\vec p,\vec q)& =& \sum_{ i = 1}^ 8 \phi_{
tT}^ {(i)} (- \vec p,\vec q) \; O_i ( \vec \sigma_{(1)}, \vec \sigma_{(3)},
\vec \sigma_{(2)}, - \vec p,\vec q) P_{23}^ s | \chi^ m \rangle \cr 
&= & (-)^ t \sum_{ i = 1}^ 8\phi_{ tT}^ {(i)} ( \vec p,\vec q)  \;
O_i ( \vec \sigma_{(1)}, \vec \sigma_{(2)}, \vec \sigma_{(3)}, \vec p,\vec q)|
\chi^ m \rangle.
\label{eq:83}
\end{eqnarray}

\noindent
The operators $ O_2, O_3, O_7 $ and $ O_8 $ are odd under exchange of 
particles 2
and 3, whereas  
$ O_1, O_4, O_5 $ and $ O_6 $ are even.

\noindent
Furthermore,  $P_{23}^ s | \chi^ m \rangle = - | \chi^ m \rangle $ leading to
\begin{eqnarray}
\lefteqn{\sum_{ i = 1}^ 8 \phi_{ tT}^ {(i)} (- \vec p,\vec q) O_i (
\vec \sigma_{(1)}, \vec \sigma_{(3)}, \vec \sigma_{(2)}, - \vec p,\vec q)
P_{23}^ s | \chi^ m \rangle  =} \cr
& &  (-)^ {t+1}  \sum_{ i = 1}^ 8\phi_{
tT}^ {(i)} ( \vec p,\vec q) O_i ( \vec \sigma_{(1)}, \vec \sigma_{(2)},
\vec \sigma_{(3)}, \vec p,\vec q)| \chi^ m \rangle.
\label{eq:84}
\end{eqnarray}
Now we use the symmetry properties of the operators $O_i$ and get
\begin{eqnarray}
\lefteqn{ \Big(\phi_{ tT}^ {(1)} (- \vec p,\vec q) O_1 - \phi_{ tT}^ {(2)}
(- \vec p,\vec q) O_2 - \phi_{ tT}^ {(3)} (- \vec p,\vec q) O_3 +
\phi_{ tT}^ {(4)} (- \vec p,\vec q) O_4 +}\cr 
& &  \phi_{ tT}^
{(5)} (- \vec p,\vec q) O_5 + \phi_{ tT}^ {(6)} (- \vec p,\vec q)
O_6 - \phi_{ tT}^ {(7)} (- \vec p,\vec q) O_7 - \phi_{ tT}^ {(8)}
(- \vec p,\vec q) O_8 \Big) | \chi^ m \rangle \cr 
&  = &  (-)^ {t+1}  \sum_{ i
= 1}^ 8\phi_{ tT}^ {(i)} ( \vec p,\vec q) O_i | \chi^ m \rangle  ~.
\end{eqnarray}
This is fulfilled if the scalar functions obey
\begin{eqnarray}
\phi_{ tT}^ {(1)} (- \vec p,\vec q) & = &  (-)^ {t+1} \phi_{ tT}^ {(1)} ( \vec
p,\vec q) \nonumber \\
\phi_{ tT}^ {(2)} (- \vec p,\vec q) & = &  (-)^ {t} \phi_{ tT}^ {(2)} ( \vec p,\vec
q) \nonumber \\
\phi_{ tT}^ {(3)} (- \vec p,\vec q) & = &  (-)^ {t} \phi_{ tT}^ {(3)} ( \vec p,\vec
q) \nonumber \\
\phi_{ tT}^ {(4)} (- \vec p,\vec q) & = &  (-)^ {t+1} \phi_{ tT}^ {(4)} ( \vec
p,\vec q) \nonumber \\
\phi_{ tT}^ {(5)} (- \vec p,\vec q) & = &  (-)^ {t+1} \phi_{ tT}^ {(5)} ( \vec
p,\vec q) \nonumber \\
\phi_{ tT}^ {(6)} (- \vec p,\vec q) & = &  (-)^ {t+1} \phi_{ tT}^ {(6)} ( \vec
p,\vec q) \nonumber \\
\phi_{ tT}^ {(7)} (- \vec p,\vec q) & = &  (-)^ {t} \phi_{ tT}^ {(7)} ( \vec p,\vec
q) \nonumber \\
\phi_{ tT}^ {(8)} (- \vec p,\vec q) & = &  (-)^ {t} \phi_{ tT}^ {(8)} ( \vec p,\vec
q) ~.
\label{eq:94} 
\end{eqnarray}

\noindent
Our standard manner \cite{stadler} to solve the 3N bound state
Faddeev equation is by iteration and using a Lanczos type
algorithm, which requires a choice  of initial amplitudes 
$\phi_{ tT}^ {(i)} ( \vec p,\vec q)$. In order to guarantee the Fermi
character for the 3N system, the symmetry requirements given in
Eq.~(\ref{eq:94}) have to be imposed for the initial amplitudes. 
It is
not difficult to explicitely demonstrate that the coupled set 
of equations (\ref{eq:79})  
conserves this symmetry requirement during the iteration provided the
numerics is reliably under control. In Ref.~\cite{fach04} it is
documented that the $  \hat {\vec  p} \cdot  \hat {\vec q}$ dependence of the
corresponding scalar functions for the full $^3$He state is rather
weak. This suggests to use a Legendre expansion for that angular
dependence, which would directly guarantee the symmetry
requirements.

It might be advantageous to replace the operators $ O_i$ by the
second form of expansion  given in Eq.~(\ref{eq:38}) which  is
directly related to a partial wave expansion. Ref.~\cite{fach04}
documents numerically that only very few terms
dominate this expansion. It  can be easily checked also for the
Faddeev amplitudes which of the two expansions are more efficient,
since solutions for realistic forces are  available in partial
wave form and from there  the scalar functions $ \phi( \vec p,
\vec q)$ and $ \tilde \phi( \vec p, \vec q)$ can be generated as
laid out in Ref.~\cite{fach04}.

\section{Summary and Outlook}
\label{sec_sum}

    Using an expansion of the Faddeev amplitudes in products of
    scalar functions $\phi_{tT}^{(i)}( \vec p, \vec q)$ with
    products of scalar expressions in spin and momenta as well as
    corresponding expansions of the NN $t$-operator and 3N forces, 
the Faddeev equations for the 3N bound state can be reformulated into 
a strictly finite
    set of coupled equations for the amplitudes $\phi_{tT}^{(i)}( \vec p, \vec
    q)$, which only depend on three variables. The single
    corresponding equation for three bosons has been solved
reliably \cite{schadow,liu2}.
For three nucleons this turns now into a coupled set of
    equations, which can be solved with modern computing resources.

     In case of 3N scattering one standard form of Faddeev
     equations is \cite{3nscatt}
\begin{eqnarray}
T|\Phi \rangle & =& tP|\Phi \rangle  + t P G_0 |\Phi \rangle  \cr
 &+& (1+t G_0)V^{(1)} (1+P)|\Phi\rangle + (1+t G_0)V^{(1)} (1+P)G_0 T | \Phi
\rangle ~.
\end{eqnarray}
We expect that a generalization of the operator expansion for
the NN $t$-operator can be found for $ T|\Phi \rangle $ as well. 
However, like for 
the system of three boson,  
these scalar functions will depend on 5 variables~\cite{liu3}:
$|\vec p|$, $|\vec q|$, $\hat{{\vec {p}}}\cdot\hat{{\vec {q}}}_{0}$, 
$\hat{{\vec{q}}}\cdot\hat{{\vec{q}}}_{0}$,
and $\widehat{({\vec{q}}_{0}\times{\vec{q}})}
\cdot\widehat{({\vec{q}}_{0}\times{\vec{p}})}$, 
where  $
\vec q_0$ is  the initial projectile momentum. The numerical
treatment is thus more demanding with respect to
interpolations compared to the 3N bound state. But again, this  
has been already 
mastered~\cite{liu3,Lin:2008sy} for
3 boson scattering  well into the GeV region.

\section*{Acknowledgments}
This work was supported by the Polish 2008-2011 science funds as a
 research project No. N N202 077435 and in part under the
auspices of the U.~S.  Department of Energy, Office of
Nuclear Physics under contract No. DE-FG02-93ER40756
with Ohio University.  
It was also partially supported by the Helmholtz
Association through funds provided to the virtual institute ``Spin
and strong QCD''(VH-VI-231).

\appendix
\section{A selected scalar functions}
\label{a1}

Below we provide expressions for 2N scalar functions
$A_{ k k'} ( p)$ and $B_{ kjk''} ( \vec
 p,\vec p')$:

\begin{eqnarray}
A_{ 1 1'} ( p)  =  3
\end{eqnarray}
\begin{eqnarray}
A_{ 1 2'} ( p)  = A_{ 2 1'} ( p)  =  0
\end{eqnarray}
\begin{eqnarray}
A_{ 2 2'} ( p)  =  \frac83 \, {\vec p}^{\, 4}
\end{eqnarray}

\begin{eqnarray}
B_{ 1 1 1} ( \vec p, \vec p\,')  =  3
\end{eqnarray}
\begin{eqnarray}
B_{ 2 1 1} ( \vec p, \vec p\,')  = B_{ 1 2 1} ( \vec p, \vec p\,')  =
B_{ 2 2 1} ( \vec p, \vec p\,')  =  B_{ 1 1 2} ( \vec p, \vec p\,')  =
B_{ 1 2 2} ( \vec p, \vec p\,')  = 0
\end{eqnarray}
\begin{eqnarray}
B_{ 1 3 1} ( \vec p, \vec p\,')  =
\left( {\vec p} \times { \vec p\,'} \, \right)^2
\end{eqnarray}
\begin{eqnarray}
B_{ 1 4 1} ( \vec p, \vec p\,')  =
\left( {\vec p} + { \vec p}^{\, \prime} \, \right)^2
\end{eqnarray}
\begin{eqnarray}
B_{ 1 5 1} ( \vec p, \vec p\,')  =
\left( {\vec p} - { \vec p}^{\, \prime} \, \right)^2
\end{eqnarray}
\begin{eqnarray}
B_{ 1 6 1} ( \vec p, \vec p\,')  =
-2 \, \left( {\vec p \prime}^{\, 2} - { \vec p}^{\, 2} \, \right)
\end{eqnarray}
\begin{eqnarray}
 B_{231}( \vec p, \vec p\,') = - p^2 ( p^2 p'^2 - (\vec p \cdot \vec p\,')^2)
- \frac {1} {3} p^2 ( \vec p \times \vec p\,')^2
\end{eqnarray}
\begin{eqnarray}
  B_{241}( \vec p, \vec p\,') = 3 ( p^2
+ \vec p \cdot \vec p\,')^2 - (\vec p \times \vec
  p\,')^2
- \frac {1} {3} p^2 ( \vec p + \vec p\,')^2
\end{eqnarray}
\begin{eqnarray}
  B_{251}( \vec p, \vec p\,') = 3 ( p^2 - \vec p \cdot \vec p\,')^2
- ( \vec p \times \vec
  p\,')^2
-\frac {1} {3} p^2 ( \vec p - \vec p\,')^2
\end{eqnarray}
\begin{eqnarray}
  B_{261}( \vec p, \vec p\,') = 6 \vec p \cdot ( \vec p
- \vec{p}\,') \vec p \cdot (\vec p +
  \vec p\,') + 2 (\vec p \times \vec p\,')^2 - \frac {2} {3} p^2 ( p^2- p'^2)
\end{eqnarray}
Expression for the scalar functions $B_{1j2}$ are obtained from $B_{2j1}$
replacing $\vec{p} \leftrightarrow \vec{p'}$.

\begin{eqnarray}
B_{ 2 1 2} ( \vec p, \vec p\,')  = 3
 ( {\vec p} \cdot { \vec p}^{\, \prime})^2  \, -
\left( {\vec p} \times { \vec p}^{\, \prime} \, \right)^2
- \frac {1} {3} p^2 p'^2
\end{eqnarray}
\begin{eqnarray}
B_{ 2 2 2} ( \vec p, \vec p\,')  = -8 \, i \,
 {\vec p} \cdot { \vec p}^{\, \prime} \,
\left( {\vec p} \times { \vec p}^{\, \prime} \, \right)^2
\end{eqnarray}
\begin{eqnarray}
 B_{232}( \vec p, \vec p\,') &=&  \frac {1}{9} (-27 (  \vec p \times  \vec p\,')^4
+p^2 p'^2 (  \vec p \times  \vec p\,')^2 
+3p'^2 ( \vec p \times (\vec p \times  \vec p\,'))^2 \cr
&+&3p^2 ( \vec p\,' \times (\vec p\,' \times  \vec p))^2
+9 (\vec p \cdot \vec p\,')^2 (  \vec p \times  \vec p\,')^2)
\end{eqnarray}
\begin{eqnarray}
 B_{242}( \vec p, \vec p\,') &=&  \frac {10} {9} p^2p'^2 ( \vec p + \vec
 p\,')^2 -p^2( p'^2 + \vec p \cdot \vec p\,')^2
- p'^2( p^2 + \vec p \cdot \vec p\,')^2 \cr
 &+& \frac {1} {3} (p^2 +p'^2) (  \vec p \times  \vec p\,')^2
\end{eqnarray}
\begin{eqnarray}
 B_{252}( \vec p, \vec p\,') &=&  \frac {10} {9} p^2p'^2 ( \vec p - \vec
 p\,')^2 -p^2( p'^2 - \vec p \cdot \vec p\,')^2
- p'^2( p^2 - \vec p \cdot \vec p\,')^2 \cr
 &+& \frac {1} {3} (p^2 +p'^2) (  \vec p \times  \vec p\,')^2
\end{eqnarray}
\begin{eqnarray}
 B_{262}( \vec p, \vec p\,') = 2( p^2 - p'^2) ( \frac {1} {9} p^2p'^2
 -  ( \vec p \cdot  \vec p\,' )^2
 + \frac {1} {3} (  \vec p \times  \vec p\,')^2)
\end{eqnarray}

The 3N scalar functions $C_{ki} ( \vec p \vec q, \vec p \vec q )$ are
symmetrical in  $k$ and $i$ indices. The diagonal values are:
\begin{eqnarray}
C_{11} ( \vec p \vec q, \vec p \vec q) &=& 2 
\end{eqnarray}
\begin{eqnarray}
C_{22} ( \vec p \vec q, \vec p \vec q) &=& 6 
\end{eqnarray}
\begin{eqnarray}
C_{33} ( \vec p \vec q, \vec p \vec q) &=& 
C_{44} ( \vec p \vec q, \vec p \vec q) = 2 (\hat{\vec p} \times \hat
{\vec q}\,)^2
\end{eqnarray}
\begin{eqnarray}
C_{55} ( \vec p \vec q, \vec p \vec q) &=& 
C_{66} ( \vec p \vec q, \vec p \vec q) = 2 \hat {\vec p}^{\,2} \hat {\vec q}^{\,2}
\end{eqnarray}
\begin{eqnarray}
C_{77} ( \vec p \vec q, \vec p \vec q) &=& 2 \hat {\vec p}^{\,2} \hat {\vec p}^{\,2}
\end{eqnarray}
\begin{eqnarray}
C_{88} ( \vec p \vec q, \vec p \vec q) &=& 2 \hat {\vec q}^{\,2} \hat {\vec q}^{\,2}
\end{eqnarray}

The nonvanishing nondiagonal values are:
\begin{eqnarray}
C_{25} ( \vec p \vec q, \vec p \vec q) &=& 
C_{26} ( \vec p \vec q, \vec p \vec q) = 
2 \hat{\vec p} \cdot \hat {\vec q}
\end{eqnarray}
\begin{eqnarray}
C_{27} ( \vec p \vec q, \vec p \vec q) &=& 
2 \hat{\vec p}^{\,2}
\end{eqnarray}
\begin{eqnarray}
C_{28} ( \vec p \vec q, \vec p \vec q) &=& 
2 \hat{\vec q}^{\,2}
\end{eqnarray}
\begin{eqnarray}
C_{56} ( \vec p \vec q, \vec p \vec q) &=& 
C_{78} ( \vec p \vec q, \vec p \vec q) = 
2 (\hat{\vec p} \cdot \hat {\vec q}\,)^2
\end{eqnarray}
\begin{eqnarray}
C_{57} ( \vec p \vec q, \vec p \vec q) &=& 
C_{67} ( \vec p \vec q, \vec p \vec q) = 
2 \hat{\vec p}^{\,2} \hat{\vec p} \cdot \hat {\vec q}
\end{eqnarray}
\begin{eqnarray}
C_{58} ( \vec p \vec q, \vec p \vec q) &=& 
C_{68} ( \vec p \vec q, \vec p \vec q) = 
2 \hat{\vec q}^{\,2} \hat{\vec p} \cdot \hat {\vec q}
\end{eqnarray}

Some other arbitrarily chosen 3N scalar functions 
$D_{kjk'}( \vec p \vec q, \vec p\,' \vec q\,', \vec p\,'' \vec q\,'')$, \newline
$D'_{kjk'}( \vec p \vec q, \vec p\,' \vec q\,', \vec p\,'' \vec q\,'')$
 and 
$H_{kjlk'} ( \vec p \vec q, \vec p \vec p\,', \vec p\,' \vec q \vec p\,''
\vec q\,'', \vec p\,''' \vec q\,'')$ 
are:
\begin{eqnarray}
D_{211}( \vec p \vec q , \vec p\,' \vec q\,', \vec p\,'' \vec q\,'') &=& -3 \\
D_{222}( \vec p \vec q , \vec p\,' \vec q\,', \vec p\,'' \vec q\,'') &=& D_{243}( \vec p \vec q , \vec p\,' \vec q\,', \vec p\,'' \vec q\,'') = 0 \\
D'_{211}( \vec p \vec q , \vec p\,' \vec q\,', \vec p\,'' \vec q\,'') &=& -3 \\
D'_{222}( \vec p \vec q , \vec p\,' \vec q\,', \vec p\,'' \vec q\,'') &=& 0 \\
H_{ 2223} ( \vec p \vec q, \vec p \vec p\,', \vec p\,' \vec q \vec p\,''
\vec q\,'', \vec p\,''' \vec q\,'') &=&0 \\
H_{ 6528} ( \vec p \vec q, \vec p \vec p\,', \vec p\,' \vec q \vec p\,''
\vec q\,'', \vec p\,''' \vec q\,'') &=& 
  i \hat{\vec p} \cdot (\vec p -\vec p\,') \times  (\vec p\,''' - \vec
  q\,'') \cr
  &-& \frac {1} {2} i ( \hat {\vec p} \times ( (\vec p -\vec p\,') \times  \vec
  p\,''')) \cdot ((\vec p -\vec p\,') \times  \vec q\,'') \cr
  &-& \frac {1} {2} i ( \hat {\vec p} \times ( (\vec p -\vec p\,') \times  \vec
  q\,'')) \cdot ((\vec p -\vec p\,') \times  \vec p\,''')
\end{eqnarray}

\end{document}